\documentclass[twocolumn,showpacs,preprintnumbers,superscriptaddress,amsmath,amssymb]{revtex4}
\usepackage{epsfig}
\usepackage{subfigure}
\usepackage{amsmath}
\usepackage{amssymb}
\usepackage{graphicx}
\usepackage{dcolumn}
\usepackage{bm}
\input epsf

\begin{document}

\title{Desynchronization waves and localized instabilities in oscillator arrays}

\author{Juan G. Restrepo}
\email{juanga@math.umd.edu}
\affiliation{
Institute for Research in Electronics and Applied Physics, 
University of Maryland, College
Park, Maryland 20742
}
\affiliation{
Department of Mathematics,
University of Maryland, College Park, Maryland 20742
}

\author{Edward Ott}
\affiliation{
Institute for Research in Electronics and Applied Physics, 
University of Maryland, College
Park, Maryland 20742
}
\affiliation{
Department of Physics and Department of Electrical and Computer Engineering,
University of Maryland, College Park, Maryland 20742
}

\author{Brian R. Hunt}
\affiliation{
Department of Mathematics,
University of Maryland, College Park, Maryland 20742
}
\affiliation{
Institute for Physical Science and Technology,
University of Maryland, College Park, Maryland 20742
}

\date{\today}

\begin{abstract}
We consider a ring of identical or near identical coupled periodic oscillators in which the connections 
have randomly heterogeneous strength. We use the 
master stability function method to determine the possible patterns at the desynchronization transition that occurs 
as the coupling strengths are 
increased. We demonstrate Anderson localization of the modes of instability, and show that such localized 
instability generates waves of desynchronization that spread to the whole array. Similar results should apply
to other networks with regular topology and heterogeneous connection strengths.   
\end{abstract}


\maketitle

In this Letter we discuss the synchronization of a large number of near-identical oscillators 
that are locally coupled with connections of random strength.
Synchronization in networks of coupled oscillators has recently received considerable interest \cite{pikovsky}, 
and has
relevance in fields like biology \cite{mosekilde}, chemistry \cite{hudson}, lasers \cite{roy,jordi}, 
and communications \cite{cuomo}.
Usually, the networks studied have been assumed to have connections of equal strength. 
In practice, the connections between different 
oscillators may have different strengths, and in some cases this strength 
could have a large spread (e.g., in biological systems). 
A model and analysis method has been proposed by Pecora and Carroll \cite{pecora2} 
to systematically determine the stability 
of the synchronized state in a network of identical coupled oscillators. 
This method, the {\it master stability function}, 
has been used to study the synchronization properties of different networks \cite{pecora3,takashi}. 
Deng et al. \cite{deng} have obtained, using the master stability function technique, 
conditions for the distribution of 
the connection strengths that yield average stability 
of the synchronized state. Galias and Ogorzalek \cite{galias} have studied the effect of 
adding small perturbations to the coupling strengths in 
relatively small arrays of coupled chaotic oscillators. 
Denker et. al. \cite{denker} have studied the effect of small coupling strength heterogeneity in networks of 
pulse-coupled oscillators.  
Our approach in this Letter will be different: we consider the coupling strengths 
to have a relatively large spread, and will discuss phenomena that can be expected when a large number 
of periodic oscillators are coupled in such a network. 
In particular, we will see that as the coupling strength is increased, 
the oscillators desynchronize in a localized region.
The localization results because the connection matrix has random 
components and the eigenvectors of this matrix are
Anderson localized \cite{anderson, lifshits}.
The effect of the localized instability spreads as a wave throughout the array, eventually
resulting in an ordered state. Remarkably, in the case where the oscillators are not identical
the final state of the locally unstable system is more ordered than in the case where the system is stable.

We consider a model system of $N$ identical dynamical units, each
one of which, when isolated, satisfies $\dot{X}_{i}  = F(X_{i})$, where $i =
1,2,\dots N$, and $X_{i}$ is the $d$-dimensional state vector for unit $i$.
(The case of nearly identical units is considered at the end of this Letter. See also \cite{ours}.) 
The oscillators, when coupled, are taken to satisfy (e.g., \cite{pecora2}) 
\begin{equation}\label{eq:coupled}
\dot{X}_{i} = F(X_{i}) - g \sum_{j = 1}^{N}
G_{ij}H(X_{j}),
\end{equation}
where the coupling function $H$ is
independent of $i$ and $j$, and the matrix $G$ is a symmetric Laplacian matrix
($\sum_{j} G_{ij} = 0$) describing the network
connections. The constant $g$ determines the global strength of the coupling.

There is an exactly synchronized
solution of Eqs.~(\ref{eq:coupled}), $X_{1} = X_{2} = \dots = X_{N} = s(t)$, whose time evolution is the same
as the uncoupled dynamics of a single unit, $\dot{s} = F(s)$.
In this Letter we will be concerned with the case where the 
synchronized state is {\it periodic}, $s(t + T) = s(t)$. 
The stability of the synchronized state can be determined from
the variational equations obtained by considering an infinitesimal
perturbation $\epsilon_{i}$ from the synchronous state, $X_{i}(t)
= s(t) + \epsilon_{i}(t)$,
\begin{equation}\label{eq:linearized}
\dot{\epsilon}_{i} = DF(s)\epsilon_{i} - g \sum_{j = 1}^{N}
G_{ij}DH(s)\epsilon_{j}.
\end{equation}

Let $\epsilon = [\epsilon_{1},\epsilon_{2},\dots,\epsilon_{N}]$,
and define  the $d\times N$ matrix $\eta =
[\eta_{1},\eta_{2},\dots,\eta_{N}]$ by $\epsilon = \eta L^{T}$, where
$L$ is the orthogonal matrix whose
columns are the corresponding real orthonormal eigenvectors of $G$; $G L = L \Lambda$,
$\Lambda = diag(\lambda_{1},\lambda_{2},\dots,\lambda_{N})$ where $\lambda_{k}$ is the eigenvalue of $G$ for
eigenvector $k$. Then Eqs.~(\ref{eq:linearized}) are equivalent to
\begin{equation}\label{eq:componentwise}
\dot{\eta}_{k} = \left(DF(s) - g \lambda_{k} DH(s)\right)\eta_{k}.
\end{equation}
The quantity $\eta_{k}$ is the weight of the $k^{th}$ eigenvector
of $G$ in the perturbation $\epsilon$.  The linear stability of
each `spatial' mode $k$ is determined by the stability of the zero
solution of (\ref{eq:componentwise}).
By introducing a scalar variable $\alpha = g\lambda_{k}$, the set
of equations given by (\ref{eq:componentwise}) can be encapsulated
in the single equation,
\begin{equation}\label{eq:master}
\dot{\eta} = \left(DF(s) - \alpha DH(s)\right)\eta.
\end{equation}
The {\it master stability  function} $\Psi (\alpha)$ \cite{pecora2} is the
largest Lyapunov exponent for this equation. This function depends only on the
coupling function $H$ and the chaotic dynamics of an individual
uncoupled element, but not on the network connectivity. The
network connectivity determines the eigenvalues $\lambda_{k}$
(independent of details of the dynamics of the chaotic units). 
The stability of the
synchronized  state of the network is determined by $\Psi_{*} =
\sup_{k} \Psi(g \lambda_{k})$, where $\Psi_{*}>0$ indicates
instability. 

\begin{figure}[h]
\begin{center}
\epsfig{file = 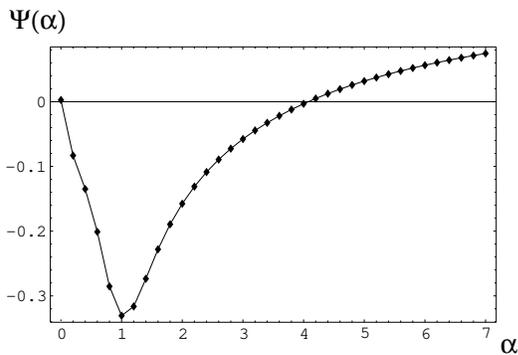, clip =  ,width=0.8\linewidth }
\caption{Master stability function $\Psi(\alpha)$ versus $\alpha$ for Eqs.~(\ref{eq:rossler}). 
At $\alpha \approx 4.15$, the master stability function becomes positive.} 
\label{fig:perio}
\end{center}
\end{figure}

As an illustrative example, we consider periodic R\"ossler oscillators, obeying the equations
\begin{eqnarray}\label{eq:rossler}
\dot{x} = - (y+z),\\\nonumber
\dot{y} = x + 0.2 y,\\\nonumber
\dot{z} = 0.2 + z(x-2.5).\\\nonumber
\end{eqnarray}
In terms of our previous notation, $d = 3$, and $X = [x,y,z]^{T}$.
The master stability function for this system is shown in Fig.~\ref{fig:perio}.
As seen in this figure, $\Psi(\alpha)$ approaches zero from negative values as $\alpha \to 0^+$. 
This is a general feature for systems where the individual, uncoupled units are stable limit cycle 
oscillators. We also see that $\Psi(\alpha)$ crosses from negative (stable) 
values to positive (unstable) values at a critical $\alpha$ value ($\alpha \approx 4.15$). 
The existence of such a 
transition is a robust feature that depends on the type of coupling and oscillator.
We now consider a network of $N$ of these oscillators nearest-neighbor coupled in a ring, such that 
the strength of each individual link is random. The coupling strengths are obtained from an
independent and identically distributed random sequence $\{a_{i}\}_{i=1}^N$. The matrix $G$ is then
\begin{equation}\label{eq:laplacian}
G = \left( \begin{array}{cccccccc}
     b_1 & -a_{1} & 0 & 0 & \cdots  & 0 & -a_{N} \\
     -a_{1} & b_2 & -a_2 & 0 & \cdots & 0 & 0  \\
     0 & -a_2 & b_3 & -a_3 & \cdots  & 0 & 0  \\
     \vdots&\vdots&\vdots&\vdots&\vdots&\vdots&\vdots\\
     -a_N & 0 & 0 & 0 & 0 & -a_{N-1} & b_N
\end{array} \right),
\end{equation}
where $b_{i} = (a_{i-1}+a_{i})$ for $i = 1,\dots, N$ (we take $a_0 \equiv a_N$).

The eigenvectors of the matrix $G$ determine the possible desynchronization patterns. 
It is known that the eigenvectors of certain types of random matrices are exponentially localized 
(e.g., Anderson localization \cite{anderson,lifshits}). In our case, the eigenvector $\{u_{i}\}_{i=1}^N$ with
eigenvalue $\lambda$ satisfies 
\begin{equation}\label{eq:evec}
t_{i+1} = a_{i+1}^{-1}(\lambda + a_i + a_{i+1} -a_i t_{i}^{-1}),
\end{equation}
where $t_{i} \equiv \frac{u_{i}}{u_{i-1}}$. Viewing Eq.~(\ref{eq:evec}) as a random
dynamical system for $t_{i}$, we find numerically that in our case,
\begin{equation}\label{eq:lyapu}
\gamma = \lim_{n \to \infty} \frac{1}{n}\sum_{i = 0}^n \log(\left| t_{i} \right| ).
\end{equation}
exists and is independent of the initial condition and noise realization. 
Eigenvectors of (\ref{eq:laplacian})
tend to have a localized amplitude peak at some location $i_{0}$ and decay like 
$\left|u_{i}\right| \propto e^{\gamma \left|i-i_{0}\right|}$ away from the peak \cite{lifshits}; $\gamma^{-1}$ 
is thus the localization length.

\begin{figure}[h]
\begin{center}
\epsfig{file = 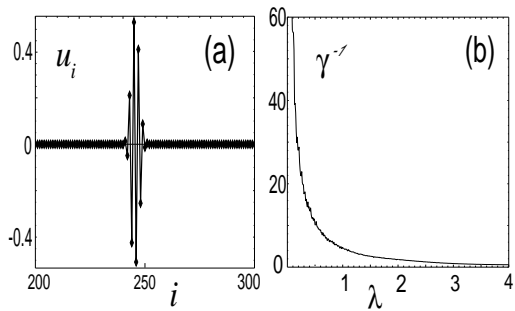, clip =  ,width= 0.8\linewidth }
\caption{(a) Eigenvector $u_{i}$ for the largest eigenvalue $\lambda = 3.61$ for a particular realization 
of the matrix $G$ in (\ref{eq:laplacian}) with $N = 500$. It is sharply localized. 
Note that the range is only $200\leq i \leq 300$.
(b) Localization length $\gamma^{-1}$ calculated using Eq.~(\ref{eq:lyapu}).} 
\label{fig:loca}
\end{center}
\end{figure}

We choose the $a_{i}$'s to be uniformly distributed in $(0.1,1)$ (note that any multiple of this
would lead to the same eigenvectors). (Since $a_{i} \geq 0.1$ we avoid the possibility $a_{i} \ll 1$ that would 
effectively disconnect the network.) The effects we will describe for this network should be regarded as an
example of what could be expected in more general networks with random coupling. 
In Fig.~\ref{fig:loca}(a) we show the eigenvector with 
largest eigenvalue for a realization of the matrix $G$ using $N = 500$. Figure~\ref{fig:loca}(b) shows
the localization length 
$\gamma^{-1}$ as a function of $\lambda$ calculated using Eq.~(\ref{eq:lyapu}).
The eigenvectors are seen to be sharply localized for the largest eigenvalues, and become less
localized as the eigenvalues decrease. 

\begin{figure}[h]
\begin{center}
\epsfig{file = 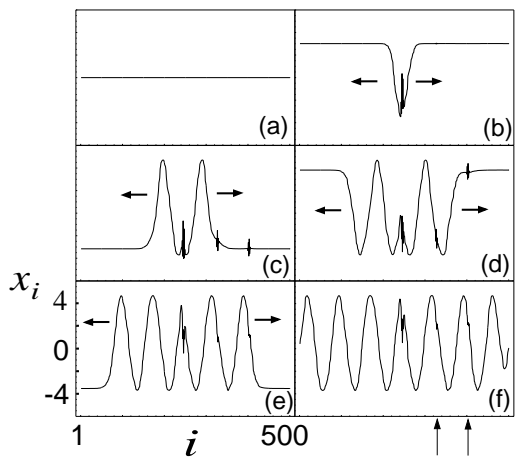, clip =  ,width=0.8\linewidth }
\caption{Plots of the $x$ coordinate of oscillator $i$ versus the site index $i$, at times (a) $0$, 
(b) $1400$, (c) $2800$, (d) $4200$, (e) $5600$, and (f) $10000$. 
All the plots have the same scale as (e).} 
\label{fig:snapshots}
\end{center}
\end{figure}

As the coupling strength $g$ is increased, the eigenvectors with largest eigenvalue
become unstable. These eigenvectors have the smallest localization length (see Fig.~\ref{fig:loca} (b)). 
We will now describe what 
occurs in this situation. We fixed the same realization of the matrix $G$ 
used in producing Fig.~\ref{fig:loca}(a). The four largest eigenvalues are 
$3.61$, $3.41$, $3.38$, and $3.30$. 
For $g = 1.24$ the eigenvector with largest eigenvalue is unstable, and the next two eigenvectors are barely unstable 
($\alpha = 4.47$, $4.23$ and $4.19$ in Fig.~(\ref{fig:perio})). 
We start with initial conditions near the 
synchronized state and then let the system evolve according to Eqs.~(\ref{eq:coupled}). In Fig.~\ref{fig:snapshots} 
we show snapshots of $x_{i}$ as a function of the site index $i$ for six successively increasing times. 
Starting from a nearly synchronized state (Fig.~\ref{fig:snapshots}(a)), the oscillators desynchronize at 
the location (see Fig.~\ref{fig:loca}(a)) of 
the localized mode (Fig.~\ref{fig:snapshots}(b)). The 
desynchronization spreads as a wave to farther regions of the array (Figs.~\ref{fig:snapshots}(c)-(e)). At the end,
the domain of the wave covers the entire array (Fig.~\ref{fig:snapshots}(f)).
This process is dominated by the most unstable mode. 
The other two less unstable modes can be seen as tiny defects  at $i\approx 327$, $402$ in the otherwise smooth wave.
(The effect of these less unstable modes is most evident in Fig.~\ref{fig:snapshots}(c). They also have a discernible,
although small, effect in the final state [arrows in Fig.~\ref{fig:snapshots}(f)].)

\begin{figure}[h]
\begin{center}
\epsfig{file = 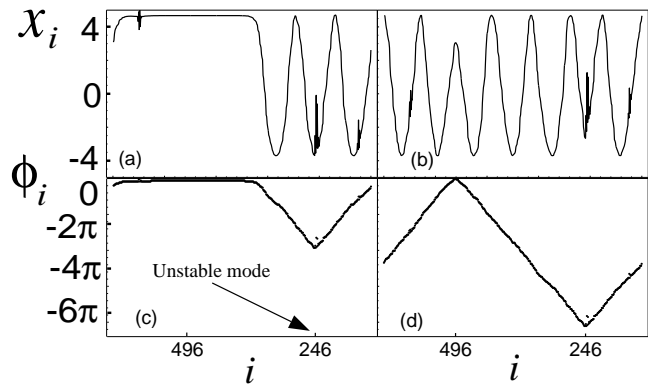, clip =  ,width=1.0\linewidth }
\caption{Plots (a) and (b) shows the $x$ coordinate of oscillator $i$ versus the site index $i$ for times $3750$ and $9660$. 
Plots (c) and (d) show the phase of oscillator $i$ at the same times as for (a) and (b) respectively. 
Compare with Eq.~(\ref{eq:target})} 
\label{fig:fase}
\end{center}
\end{figure}

The final state and the process leading to it can be understood in terms of the phase of the oscillators. Define the phase
$\phi(i,t) \equiv 2\pi\{n(i,t)+ (t - t_{-}(i,t))(t_{+}(i,t) - t_{-}(i,t))^{-1}\}$, 
where $t_{-}(i,t) = \max\{s: x_{i}(s) = 0, \dot{x}_{i} > 0, s \leq t\}$,
$t_{+}(i,t) = \min\{s: x_{i}(s) = 0, \dot{x}_{i} > 0, s > t\}$, and $n(i,t)$
is an integer chosen so that $\phi$ is a continuous function of $t$ and that 
$\phi(i+1,t)$ is close to $\phi(i,t)$ for all $i$. 
Figure \ref{fig:fase} shows
two snapshots of the $x$ coordinate and the phase as defined above as a function of $i$ 
(the $i$ origin was displaced so that
what happens opposite the location of the unstable mode can be observed clearly, 
and for each time a constant was added to $\phi$
so that $\max_{i} \phi = 0$). 
As can be observed in the Figs.~\ref{fig:fase} (a) and (c),
a region with a constant phase gradient expands on both sides of the unstable mode. 
In the final state (Figs.~\ref{fig:fase}(b) and (d)) the phase has a minimum at the location of the unstable mode 
and increases linearly on both sides reaching a maximum at the opposite end of the ring. 
This phase profile increases uniformly with time.
The cause of this phenomenon is that, as
the oscillators in the region of the unstable mode desynchronize, 
they go to limit cycles that have a slightly lower frequency 
than that of the original orbit. Oscillating at a slower pace than the others, they 
drag the adjacent oscillators, and these drag theirs in turn, continuing 
until an equilibrium is reached.
An equation describing approximately the evolution of the 
phase of the oscillator at location $\xi$ and time $t$, $\phi(\xi,t)$,
in a chain of diffusively coupled oscillators is given in the continuous limit by \cite{kuramoto}
\begin{equation}\label{eq:phaseequation}
\frac{\partial\phi}{\partial t} = a \frac{\partial^2 \phi}{\partial \xi^2} 
+ b \left(\frac{\partial \phi}{\partial \xi}\right)^2 + w(\xi),
\end{equation}
where $w(\xi)$ is the frequency of the oscillator at location $\xi$, and $a$ and $b$ are constants. 
If this frequency is sufficiently smaller (larger) in a localized region and $b$ is negative (positive),
the equation predicts the development of  
waves that emanate from that region. 
The phase profile resulting from such forcing in a small region centered at the origin
$(\left|\xi\right| < l)$
can be approximated for large $\xi$ and $t$ as \cite{kuramoto}
\begin{equation}\label{eq:target}
\phi(\xi,t) = w_{0} t - \max(0, k(v t - \left|\xi\right|)),
\end{equation}
where $w_{0} = w(\xi)$ for $\left|\xi\right| > l$
and $k$ and $v$ depend on $a$ and $b$ and $w(\xi)$. 
For appropriate $k$ and $v$, equation (\ref{eq:target}) agrees well with Figs.~\ref{fig:fase} (c) and (d).
\begin{figure}[h]
\begin{center}
\epsfig{file = 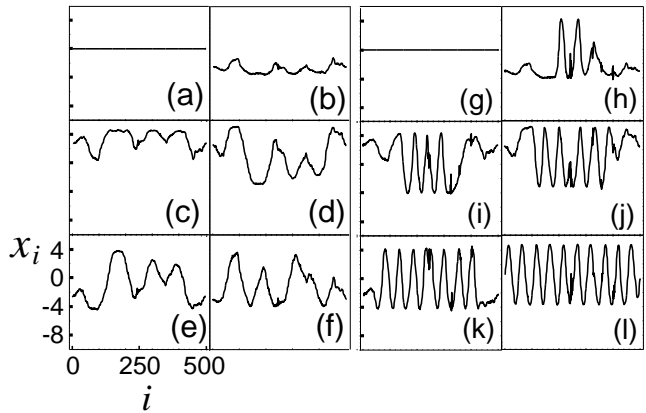, clip =  ,width=1.0\linewidth }
\caption{ Each plot shows the $x$ coordinate of oscillator $i$ as a function of the site index $i$.
The time is $0$, $1400$, $2800$,
$4200$, $5600$, and $9970$ for plots (a) to (f) and similarly for plots (g) to (l).
A parameter mismatch was introduced in the oscillators. 
(a)-(f): All the modes are stable. (g)-(l): The pattern is organized by an unstable mode as in Fig.~\ref{fig:snapshots}(f).
All the plots have the same scale as (e).} 
\label{fig:compara}
\end{center}
\end{figure}
In the example presented above, the pattern created by the unstable mode can be regarded as a more disordered
synchronization than that of the original identical synchronization. 
However, in realistic situations, an unstable mode can actually make synchronization more orderly. 
In real systems, small differences in the parameters or small noise are expected. 
Under these circumstances, the 
different oscillators will be subject to small perturbations. 
The modes with eigenvalues close to zero 
have a master stability function close to zero (see Fig.~\ref{fig:perio})
and also are nearly unlocalized (see Fig.~\ref{fig:loca}(b)).
Thus, the phase of each oscillator will be
subject to perturbations whose projection onto the nearly unlocalized modes are only very weakly damped. 
The identical synchronization of the array is thus spoiled by mismatch or noise. 
As an illustration, we randomly perturb the parameters of 
the different oscillators, so that they lie within $\pm3\%$ of the original parameters. 
We then solved Eqs.~(\ref{eq:coupled}) with $g = 1.1$ and $g = 1.24$. 
For $g = 1.1$, all the modes are stable; in the case $g = 1.24$, three modes are stable as discussed above.
In Figs.~\ref{fig:compara} (a)-(f) we show snapshots of the case $g=1.1$, and in Figs.~\ref{fig:compara} (g)-(l) we show
the corresponding snapshots for the case $g = 1.24$. 
When all of the modes are stable, the system exhibits a state in which there is 
erratic slow variation of the $x_{i}$ with $i$.
When there is an unstable mode, however, a more organized state is reached.
If one picks two different 
oscillators $j$ and $k$,  
they will satisfy asymptotically $X_{j}(t - \tau) = X_{k}(t)$,
where $\tau$ is a simple function of $j$ and $k$ (see Fig.~\ref{fig:fase} (d)). 
Thus the oscillators are pairwise lag synchronized \cite{lag}.
In realistic large arrays of periodic oscillators, it might be convenient to have one unstable mode. 
This mode will, despite its localized nature, induce global organization of the system (Fig.~\ref{fig:compara}).

In conclusion, we find that large arrays of periodic oscillators locally coupled by connections 
of randomly heterogeneous strength can experience a desynchronization transition characterized by the appearance
of unstable Anderson localized modes. Furthermore, we find that, past the transition, the localized mode plays
the key role in organizing the final global pattern of the system oscillations.

Acknowledgements: We thank R.E. Prange for useful discussion. This work was sponsored by ONR (Physics) and by
NSF (contracts PHYS 0098632 and DMS 0104087).

\end{document}